\documentclass[%
preprint,
superscriptaddress,
 amsmath,amssymb,
 aps,
 prx,
floatfix,
]{revtex4-2}

\usepackage[english]{babel}
\usepackage{graphicx}% Include figure files
\usepackage{dcolumn}% Align table columns on decimal point
\usepackage{bm}% bold math
\usepackage{placeins}

\usepackage{hyperref}

\usepackage{color}
\usepackage[utf8]{inputenc}
\usepackage{float}

\begin{document}
\title{Collective effects and synchronization of demand in real-time demand response}

\author{Chengyuan Han~\href{https://orcid.org/0000-0001-5220-402X}{\includegraphics[width=3.2mm]{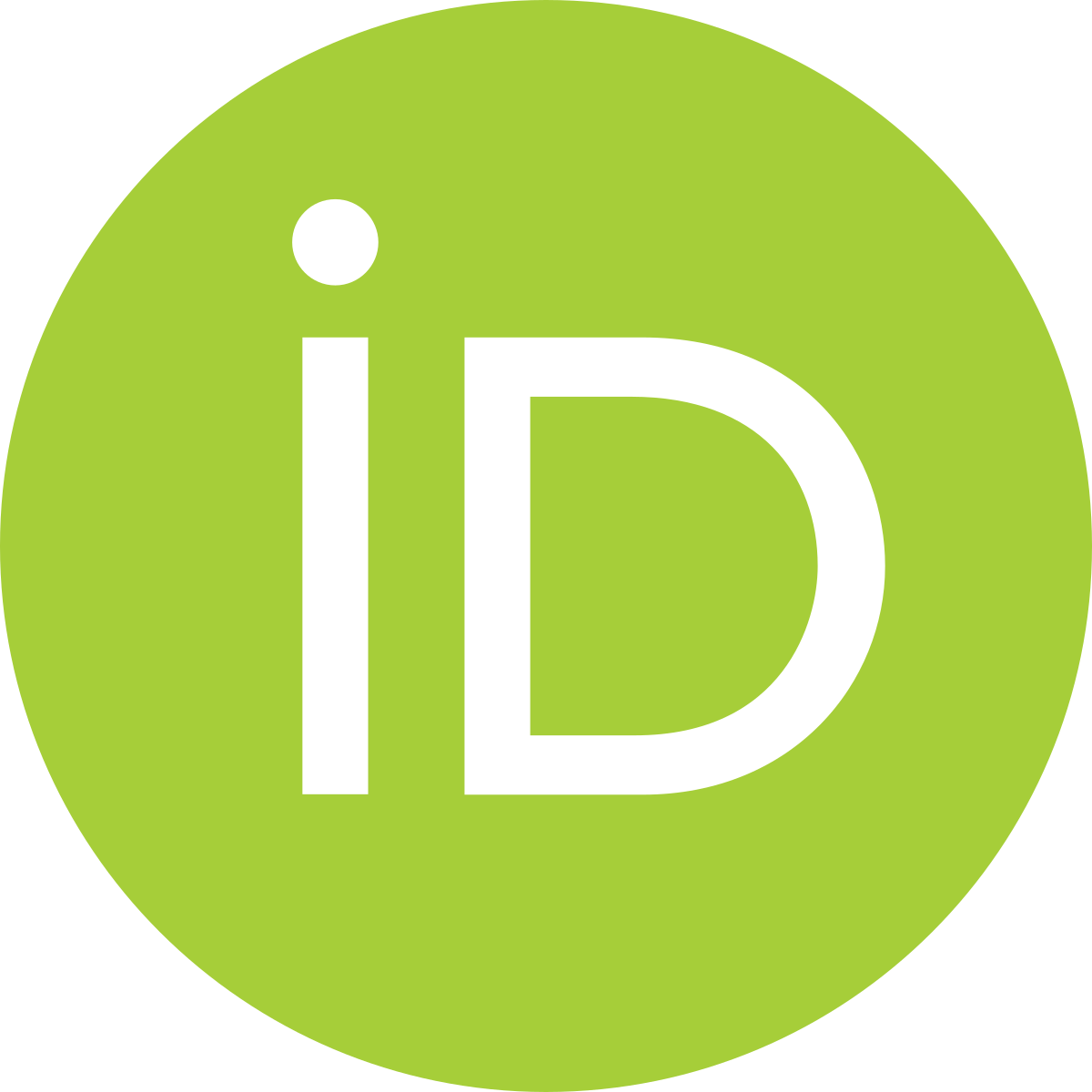}}}
 \email{ch.han@fz-juelich.de}
  \affiliation{Forschungszentrum J\"ulich, Institute for Energy and Climate Research (IEK-STE), 52428 J\"ulich, Germany}
    \affiliation{Institute for Theoretical Physics, University of Cologne, 50937 K\"oln, Germany}

\author{Dirk Witthaut~\href{https://orcid.org/0000-0002-3623-5341}{\includegraphics[width=3.2mm]{orcid.png}}}%
 \email{d.witthaut@fz-juelich.de}
  \affiliation{Forschungszentrum J\"ulich, Institute for Energy and Climate Research (IEK-STE), 52428 J\"ulich, Germany}
    \affiliation{Institute for Theoretical Physics, University of Cologne, 50937 K\"oln, Germany}

\author{Leonardo~Rydin~Gorj\~ao~\href{https://orcid.org/0000-0001-5513-0580}{\includegraphics[width=3.2mm]{orcid.png}}}
\email{leonardo.rydin@gmail.com}
\affiliation{Forschungszentrum J\"ulich, Institute for Energy and Climate Research (IEK-STE), 52428 J\"ulich, Germany}
\affiliation{Institute for Theoretical Physics, University of Cologne, 50937 K\"oln, Germany}

\author{Philipp C. Böttcher~\href{https://orcid.org/0000-0002-3240-0442}{\includegraphics[width=3.2mm]{orcid.png}}}
\email{p.boettcher@fz-juelich.de}
  \affiliation{Forschungszentrum J\"ulich, Institute for Energy and Climate Research (IEK-STE), 52428 J\"ulich, Germany}
\date{\today}

\begin{abstract}
Future energy systems will be dominated by variable renewable power generation and interconnected sectors, leading to rapidly growing complexity. Flexible elements are required to balance the variability of renewable power sources, including backup generators and storage devices, but also flexible consumers. Demand response aims to adapt the demand to the variable generation, in particular by shifting the load in time. In this article, we provide a detailed statistic analysis of the collective operation of many demand response units. We establish and simulate a model for load shifting in response to real-time electricity pricing using local storage systems. We show that demand response drives load shifting as desired but also induces strong collective effects that may threaten system stability. The load of individual households synchronizes, leading to extreme demand peaks. We provide a detailed statistical analysis of the grid load and quantify both the likeliness and extent of extreme demand peaks. 
\end{abstract}

\maketitle

\section{Introduction}

The mitigation of climate change requires a comprehensive transformation of our energy system towards renewable sources \cite{Roge15}. Wind and solar power have enormous potential \cite{jacobson2011providing} and have become fully cost-competitive in recent years \cite{irena2020}. However, system integration of renewable power sources remains a challenge as generation fluctuate on multiple time scales \cite{milan2013turbulent,Anvari16,Staffell2018,wohland2019significant}. Hence, methods of statistical physics and complexity science are becoming essential to understand the dynamics and operation of future energy systems \cite{brummitt2013transdisciplinary,timme2015focus}.

A variety of methods are being used and developed to balance the fluctuations of renewable power generation, including different storage techniques \cite{victoria2019role} and flexible balancing power plants \cite{elsner2016flexibilitatskonzepte}. Furthermore, the electricity sector may be coupled to other sectors, e.g., heating and industry, providing additional flexibility \cite{orths2019flexibility}. In addition, flexibility can be introduced on the demand side. Techniques to adapt to the fluctuating generation are commonly referred to as demand response (DR) and are heavily discussed in the literature. (see \cite{palensky2011demand,siano2014demand} for recent reviews). The adoption of DR requires financial incentives for the respective user \cite{yan2018review}. For instance, users may adapt their demand to the current electricity prices in almost real-time to reduce overall costs \cite{conejo2010real}. However, the adaption of DR at the household level is lacking behind \cite{christensen2019engage} as social and behavioral obstacles are not overcome.

In this article, we address the operation and implications of DR from a systemic statistical perspective. Without DR, the actions of single consumers, i.e., the switching of a single device, can be considered an independent stochastic event. In a large interconnected power system, demand fluctuations of individual households average out, and the total grid load varies rather smoothly. In the spirit of the central limit theorem, we can assume that the residual fluctuations of the total grid load around the smooth daily profile follow a normal distribution. This assumption is no longer valid for real-time DR, where the customer demands are adapted according to a common input signal, the electricity price, and thus are no longer independent. Collective effects may then fundamentally alter the statistics of the electricity demand.

To study the potential impacts of DR, we simulate the operation of a household DR system based on real-time pricing using a coarse-grained model and investigate the impact on the resulting electricity demand time series. On average, demand is shifted to periods of low prices as desired, but we instead focus on the statistics of the time series and collective effects emerging for many households all reacting to the same real-time price signals. It has been shown in the statistical physics community that such common inputs can fundamentally change the statistics \cite{newman1996avalanches,krause2015econophysics}.
In fact, the behaviour of different households can synchronize, which leads to heavy-tailed distributions of the aggregated demand. Events with a strongly simultaneous demand may arise, which may be adverse to power system stability. 

\section{Models and Methods}

\begin{figure}
    \centering
    \includegraphics[width=0.9\textwidth]{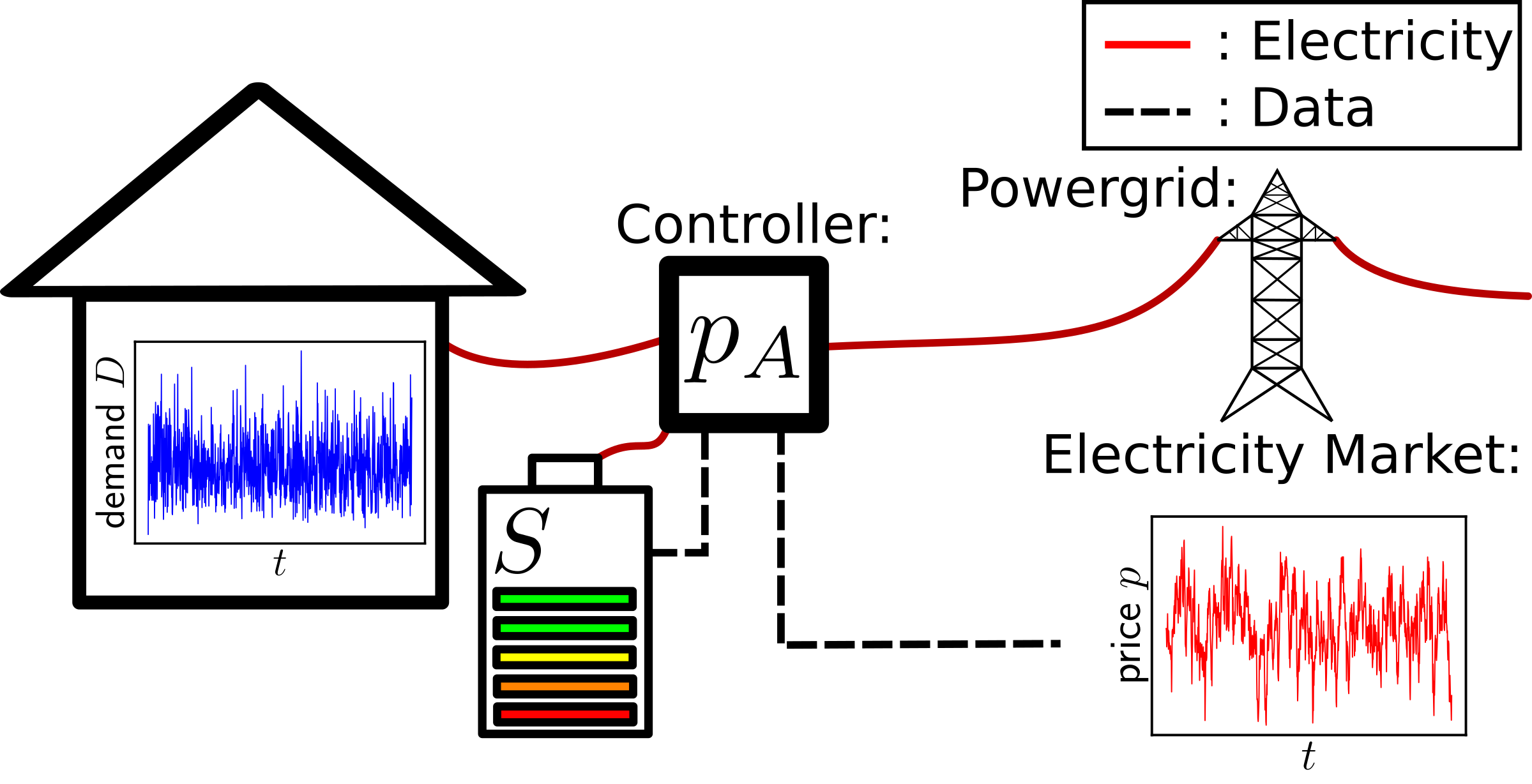}
    \caption{Sketch of the household demand response model analyzed in the present paper.
    The household demands the electric power $D(t)$ during time interval $t$.
    This demand can either be satisfied by drawing power from the grid or a local battery electric storage system (BESS) with capacity $S_{\rm Cap}$. Whether power is drawn from the grid to meet the demand and charge the BESS is decided by a controller on the basis of the real-time price $p(t)$ and the state of charge of the BESS $S(t)$. We consider an elementary control law, where $p(t)$ is compared to an acceptable price $p_a$, that is a monotonously decreasing function of $S(t)$. 
    }
    \label{fig:scheme}
\end{figure}

We consider a coarse-grained model of real-time DR. A set of $N$ households try to minimize their electricity costs by adapting their power supply, as shown in Fig.~\ref{fig:scheme}. Each household $j=1,\ldots,N$ is characterized by its residual power demand time series $D_j(t)$, which equals the final demand minus local renewable generation, e.g., by a photovoltaic source. 

The DR is realized via a small battery electric storage system (BESS) with capacity $S_{\rm Cap}$, which allows for a shifting of electricity consumption. That is, we consider only DR actions that do not require any active participation or behavioural changes by the consumer, i.e., fully automated by a controller. The residual power demand $D_j(t)$ of each household $j$ can be covered either by electricity stored in the battery or by buying electricity from the market by paying the price $p(t)$ per unit of energy. Market prices are typically updated in hourly or quarter-hourly steps. Hence, we simulate system operation in discrete time steps of length $\Delta t = $ one hour. In the following, energy is always given in units of kWh, and the power demand is given in units of kW.

The basic operation of the BESS system is determined by a controller that consider the current price and the state of the BESS. In each time interval $t$, the controller at household $j$ determines the amount of energy $E_j(t)$ purchased from the grid.
Neglecting losses, the energy stored in the BESS increases by a portion of the purchased energy and decreases by the residual energy due to demand $D_j(t) \cdot \Delta t$. The state of charge of the BESS $S_j(t)$, defined relative to the capacity $S_{\rm Cap}$, thus evolves as
\begin{equation}
    S_j(t+\Delta t) = S_j(t) + \frac{E_j(t) - D_j(t) \times \Delta t}{S_{\rm Cap}} \, .
    \label{eq:state-of-charge}
\end{equation}
The characteristics of the demand time series $D_j(t)$ and the operation of the controller that determines $E_j(t)$ are discussed below.

\subsection{Demand patterns}
\label{sec:demand-model}

This paper deals with the statistical properties of electricity consumption and collective effects emerging in the system compromised of many users, such that a large amount of input data is required.
We use a recently developed statistical model for the demand time series $D_j(t)$, which captures essential features of real-world demand fluctuation patterns \cite{anvari2020data}.

In the model, the demand time series of a household is given by a stochastic process 
\begin{align}
    D_j(t) &= \sqrt{ \sum\limits_{k=1}^{L} x_k^2(t) } + \mu_{MB}, \label{eq:mb_demand}
\end{align}
where the $x_k(t)$ are independent Ornstein-Uhlbeck (OU) processes with zero mean, diffusion constant $\sigma_{OU}$ and mean-reversion strength $\gamma_D$
\cite{gardiner2009handbook} and we choose $L=3$ following \cite{anvari2020data}. This model leads to the stationary distribution of the demand $D_j$ being described by the Maxwell-Boltzmann distribution
\begin{align}
    P(D_j) &= \frac{1}{\sigma^3_{MB}} \sqrt{\frac{2}{\pi}} (D_j - \mu_{MB})^2 \cdot \exp\left[ - \frac{(D_j - \mu_{MB})^2}{2\sigma_{MB}^2}\right].
    \label{eq:demand:OU}
\end{align}
We note that the standard deviation of the individual Ornstein-Uhlenbeck processes equals the scale parameter of the Maxwell-Boltzmann distribution, $\sigma_{MB}=\sigma_{OU}$.
Hence, distributions with different variability can be readily generated by tuning the diffusion parameter in the individual OU processes $x_k(t)$.

In the numerical simulation, we generate the individual OU processes using a Markov Chain Monte Carlo method. The transition probability from state $x_0$ at time $t_0$ to state $x_1$ at time $t_0 + \Delta t$ is given by
\begin{align}
P (x_1, t_0 + \Delta t| x_0, t_0) &= \sqrt{\frac{\gamma_D}{\pi \sigma^2 (1-e^{-2 \gamma_D \Delta t})}} \exp{\left[ - \frac{\gamma_D (x_1 - x_0 e^{- \gamma_D \Delta t} - \mu_{D} (1 - e^{-\gamma_D \Delta t}))^2}{\sigma^2 (1 - e^{-2 \gamma_D \Delta t})}\right]}. 
\label{eq:trans_OU}
\end{align}
While the authors in Ref.~\cite{anvari2020data} used the offset $\mu_{MB}$ to fit different measured load time series, our analysis only necessitates the correct stochastic behavior of the resulting load time series and thus we set $\mu_{MB}=0$ for the analysis.
Furthermore, we choose the scale parameter to give an average demand of $\langle D_j \rangle = 0.5$kW, resulting in about 12kWh consumption per household and day. The choice of the mean reversion rate $\gamma_D$ determines the time scale of the stochastic demand series. We choose this to be set to $\gamma_D=1$h$^{-1}$.

\subsection{Price time series}
\label{sec:price-model}

The real-time electricity price $p(t)$ is the essential variable driving the operation of the BESS controller. To enable more detailed stochastic investigations, we use synthetic time series, which are designed to reproduce essential statistic features of real-world time series. In particular, we model $p(t)$ as a one-dimensional OU process with parameters $\mu_p$, $\sigma_p$, and $\gamma_p$.
The mean and the standard deviation are are set to $\mu_p=39.46$~ct/kWh and $\sigma_p = 20.69$~ct/kWh, respectively, corresponding to the value observed in German intra-day spot market in the year 2014-2020.
The mean reversion rate $\gamma_p$ is kept as a tunable parameter to analyze the impact of correlations on the DR effect. We mostly use $\gamma_p=0.2$~h$^{-1}$ for illustrative purposes, but give the final results for different values of $\gamma_p$. The transition probabilities are given by an analog expression as in \eqref{eq:trans_OU}.

We note that consumer prices are generally much higher than wholesale market prices. However, any constant shift or scaling of the prices does not affect the results of our simulations. In fact, such a rescaling will only lead to an equivalent rescaling of the acceptable prices $p_{a,j}$. 

\subsection{Controller model \label{sec:controller_model}}

The BESS control system must determine how much electrical energy $E_j(t)$ is purchased from the grid in the time interval $t$. The development of optimal control algorithms for DR is a wide research field, and important progress has been made (see \cite{jordehi2019optimisation,godina2018model} for recent reviews). The scope of this study is a very different one, focusing on collective effects and emergent statistical properties. Hence, we keep the controller model as concise as possible.

First, we do not include any forecasting in the control law. Decisions are made on the basis of the current state of charge of the BESS $S_j(t)$ and the electricity price $p(t)$. In addition, the controller must take into account the variable $D_j(t)$ to ensure that the demand is always met and the battery limits are obeyed.

Second, we assume that the controller has only two basic options: Either it chooses to cover the demand completely from the battery such that $E_j(t)=0$ kWh, or it chooses to draw power from the grid to recharge and satisfy the demand. Recharging is always done at a maximum charging rate $c_r \, S_{\rm Cap}$, where $c_r \in [0,1]$ h$^{-1}$ is a tunable parameter. In this case, the household will draw the energy
\begin{equation}
    E_j(t) = (D_j(t) + c_r \, S_{\rm Cap}) \, \Delta t
    \label{eq:energy-at-charge}
\end{equation}
from the grid. Small adjustments must be made to ensure the demand is always met and the battery is never overloaded, i.e.,~$S_j(t+1) \in [0,1]$ is always satisfied. Revisiting Eq.~\eqref{eq:state-of-charge}, we find the following constraints. If the state of charge is too low to cover the demand in the current time interval, $E_j(t) = 0$ is impossible, and the BESS has to draw the energy $E_j(t) = D_j(t) \Delta t - S_j(t) S_{\rm Cap}$ from the grid. If the BESS is almost full such that Eq.~\eqref{eq:energy-at-charge} would lead to overloading, the BESS can only draw the energy  $D_j(t) \, \Delta t + S_{Cap} (1-S_j(t))$ from the grid.

Finally, we assume that the decision of whether to draw energy from the grid or not is reached by comparing the market price $p(t)$ to an acceptable price $p_{a,j}(t)$. Hence, the control law can be formulated as
\begin{align}
    E_j(t) = 
    \left\{ 
    \begin{array}{l l l}
        \min\left[ S_{Cap} (1-S_j(t)) + D_j(t) \cdot \Delta t, 
        (c_r \cdot S_{Cap} + D_j(t)) \cdot \Delta t \right]  & \; \mbox{if} \; & p(t)  <  p_{a, j}(S_j(t)) \\
        \max\left[ 0,D_j(t)\cdot\Delta t-S_j(t) \cdot S_{\rm Cap} \right]  
        &                 & p(t) \ge p_{a, j}(S_j(t)).
    \end{array}
    \right.
\end{align}

The acceptable price depends on the state of change $S_j(t)$ of the BESS. If the BESS is almost fully charged, there is no need to purchase electricity such that $p_{a,j}$ will be large. If the BESS is almost empty, recharging is urgent, and $p_{a,j}$ will be small. In the following, we assume a simple affine linear law
\begin{align}
    p_{a,j}(t) &= k + (q - k) \, S_j(t). \label{eq:linear_law}
\end{align}
Note, the parameters for $q$ and $k$ in Eq.~\eqref{eq:linear_law} give the acceptable price for a full and empty BESS, respectively. The actual value of the parameters $q$ and $k$ are determined to optimize the total costs of a single household and depend on the properties of the demand, price statistics, and the BESS itself. We will discuss this aspect in the following section.

\section{Demand Response Effect at the household level}

The DR system shifts the electricity demand of the households in time. Without DR, a household consumes the demand $D_j(t)$ directly from the grid; with DR, the purchases are instead given by the time series $E_j(t)$. By shifting to time intervals of lower prices, DR can thus reduce the total electricity cost of a household. We first analyze this effect from the perspective of a single household before we turn to systemic effects and statistical properties in the next section.
\begin{figure}
    \centering
    \includegraphics[width=\columnwidth]{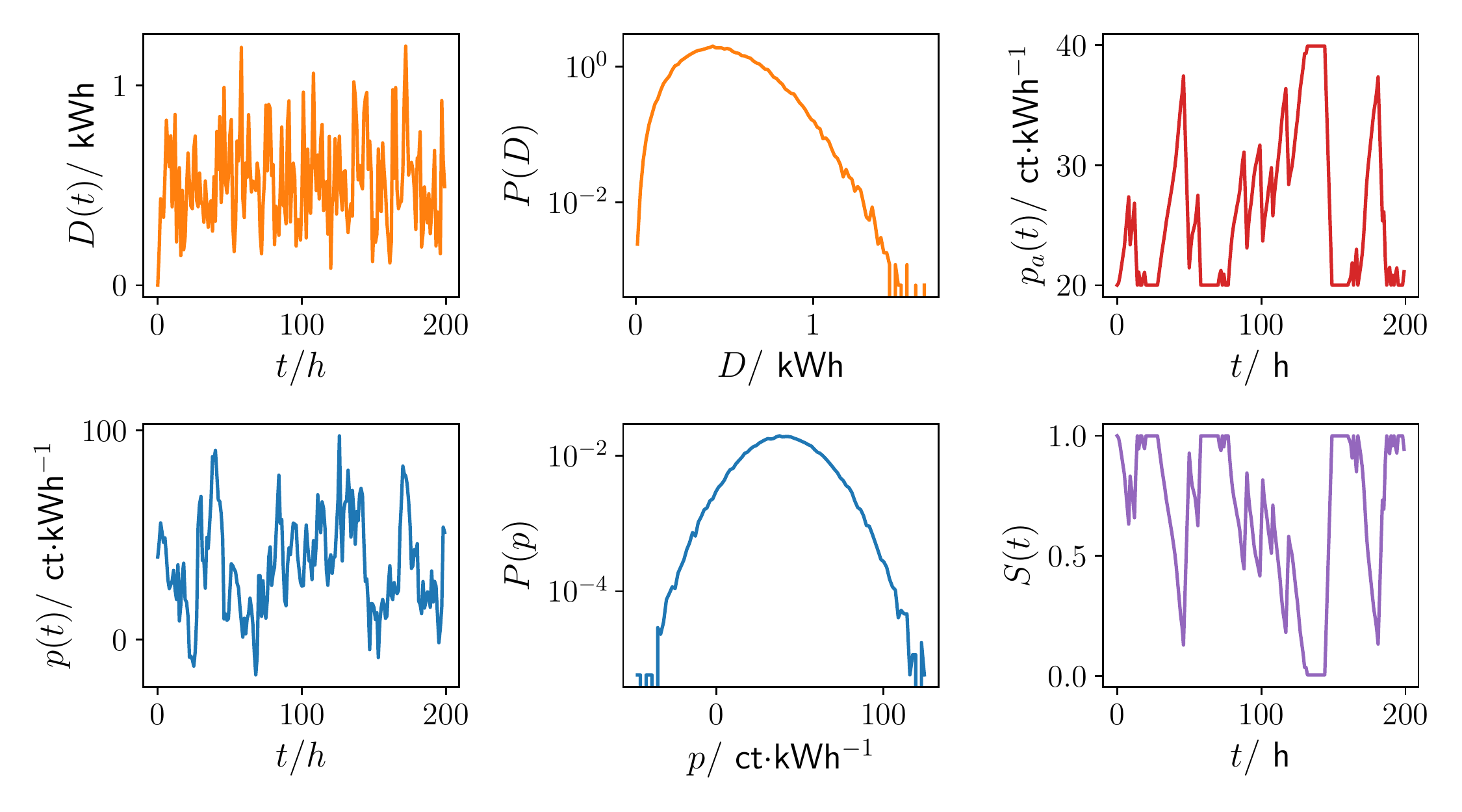}
    \caption{Sample simulation for a single household showing the dynamics. 
    The parameters defining the household were set to a battery size of $S_{\rm Cap}=10$kWh and a charging rate of $c_r=0.2$ h$^{-1}$, while the control parameters were chosen as $q=20$ ct/kWh and $k=40$ ct/kWh.
    From top left clockwise: Demand $D$, acceptable price $p_a$, price $p(t)$, and state of charge $S$. }
    \label{fig:timeseries-example}
\end{figure}
A sample simulation can be seen in Fig.~\ref{fig:timeseries-example}.
As the price is modeled as an Ornstein-Uhlenback process, its stationary distribution $P(p)$ follows a Gaussian distribution, while the demand distribution follows a Maxwell-Boltzmann distribution (left column of Fig.~\ref{fig:timeseries-example}). Using the control function described in Section~\ref{sec:controller_model}, the state of charge $S$ and acceptable price $p_a$ interact to drive the system dynamics.

\begin{figure}
    \centering
    \includegraphics[width=\columnwidth]{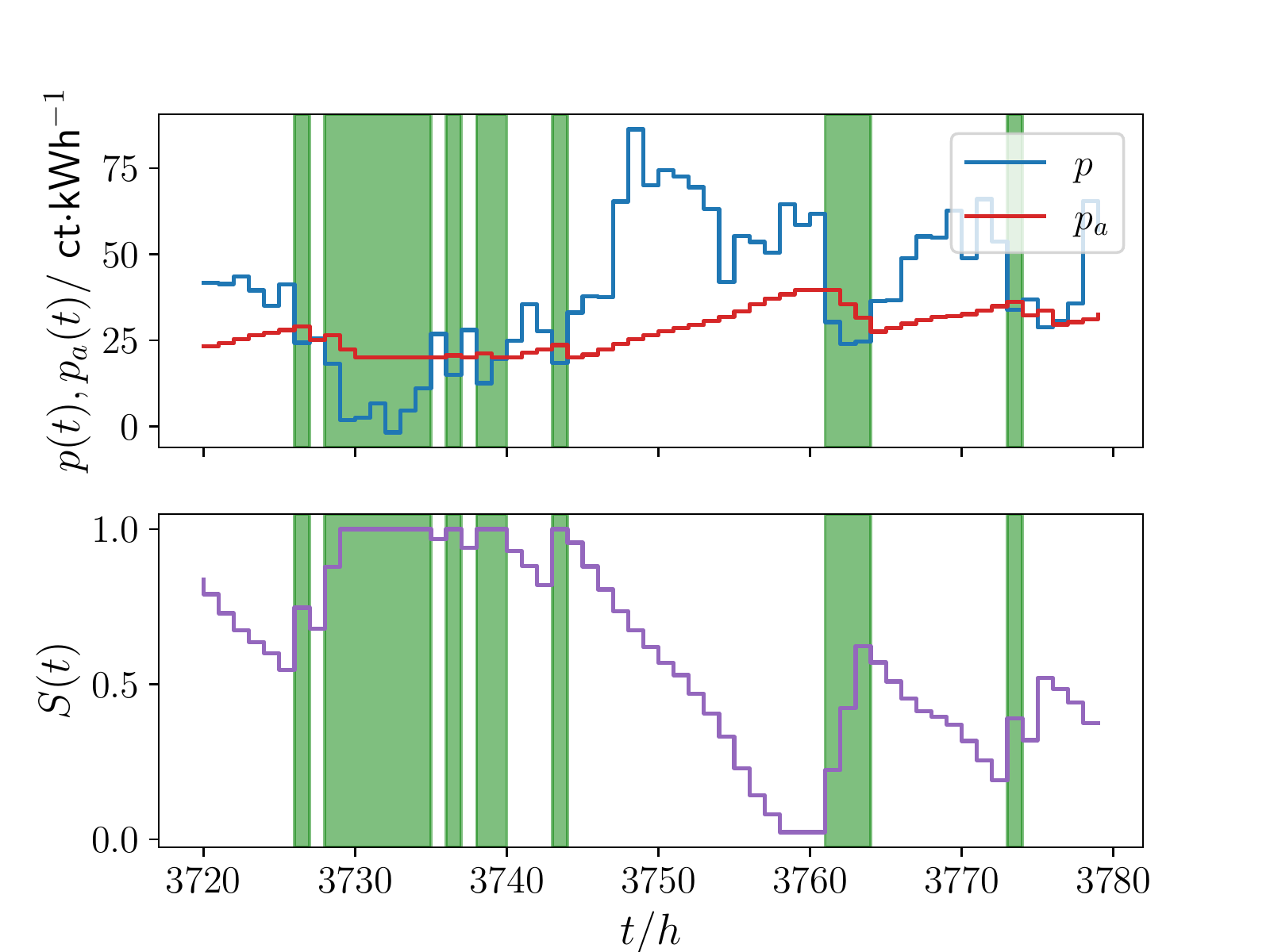}
    \caption{Example of the demand response for a single household with battery size $S_{\rm C}=10$kWh and a charging rate $c_r=0.2$ h$^{-1}$, while the control parameters were chosen as $q=20$ ct/kWh and $k=40$ ct/kWh.
    At times where the acceptable price $p_a$ is large then the price $p$ (green shaded regions), the storage is charged by $c_r \cdot S_{\rm Cap}$ if it is not full already.
    This way, the saved up energy can be used to avoid the high price regions in the middle and thus lowering the money that would have to be paid.}
    \label{fig:demand_effect_example}
\end{figure}

To understand the underlying dynamics, we need a closer look at the time evolution of the price $p$, acceptable price $p_a$, and the state of charge $S$. In Fig.~\ref{fig:demand_effect_example}, a short time window of the same simulation as in Fig.~\ref{fig:timeseries-example} is presented. The time windows where the acceptable price $p_a$ is above the market price $p(t)$, i.e.,~where the battery is charged if the limits are not exceeded, are indicated by the green shaded regions. At times where the price is too large, the battery can be used to cover the demand.
Thus, the demand has been shifted away from the times of high prices to the green shaded time regions.
To quantify the impact of the DR system for a single household, we consider the average cost that a household $j$ has to pay for the energy drawn from the grid in $N_t$ time steps,
\begin{equation}
    %\mu_{C,j} = N_t^{-1} \sum_{t=1}^{N_t} p(t) \, E_j(t),
    \mu_{C,j} = N_t^{-1} \sum_{i=1}^{N_t} p(t_i) \, E_j(t_i),
    \label{eq:def-mu_c}
\end{equation}
as well as its volatility expressed by the standard deviation $\sigma_C$. We assume that all customers individually minimize their average costs $\mu_C$ and design the controller accordingly.
Furthermore, we consider the mean $\mu$ and the standard deviation $\sigma$ of the time series $S(t)$ and $E(t)$ to characterize the operation of the DR system.
Obviously, all characteristics depend on the properties of the BESS system and the controller as well as the properties of the stochastic processes $D_j(t)$ and $p(t)$.
In the following, we fix the parameters of the stochastic processes to the values given in the previous section and focus on the BESS and control system.

\begin{figure}
    \includegraphics[width=0.9\columnwidth]{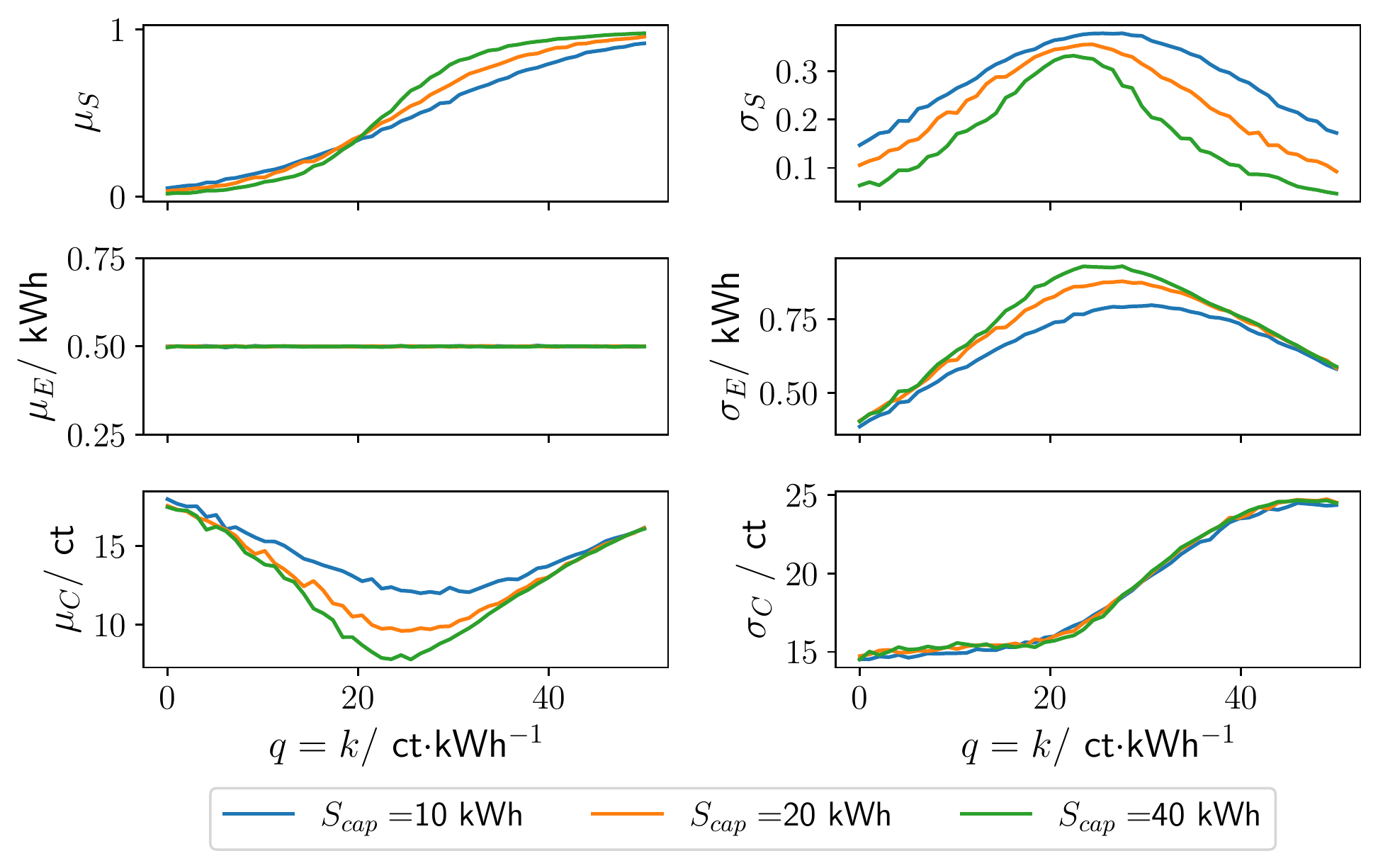}
    \caption{
    Operation of the DR/BESS system of a single household for simplified control law.     
    The panels show the mean $\mu$ and the standard deviation $\sigma$ of the state of charge $S_j(t)$ and the purchases $E_j(t)$, as well as the
    average electricity price paid by the household 
    \eqref{eq:def-mu_c} together with the volatility. The respective quantities are plotted as a function of the acceptable price 
    $p_{a,j} = k$, which is assumed to be constant here. We observe a minimum in the average price $\mu_C$ that gets more pronounced with increasing storage capacity $S_{\rm Cap}$. At the this optimum point, $\sigma_E$ and $\sigma_S$ assume a maximum. 
    \label{fig:constant_law} 
    }
\end{figure}

To begin with, we consider an even simpler control law with a constant acceptable price $p_{a,j} = k$, see Fig.~\ref{fig:constant_law}. This simplified treatment provides some fundamental insights into the operation of the BESS, which is helpful for the analysis of the full system provided below. We find that even in this simple case, a substantial reduction of the electricity costs is possible. For a large BESS with capacity $S_{\rm Cap} = 40$, we find a reduction by reduction by more than $50\%$.

In all cases, we find that there is an optimum value of the acceptable price $k^*$, for which the average electricity price $\mu_C$ assumes a minimum.
Notably, this optimum value is considerably lower than the average market price. 
For $p_{a,j} = k^*$, the system makes use of the battery in an optimum way.
It is heavily charged and discharged such that the standard deviation $\sigma_S$ assumes a maximum.
States with high and low charges are equally probable such that. The purchases $E_j(t)$ are also most volatile at the optimum point.

\begin{figure}
    \includegraphics[width=0.9\columnwidth]{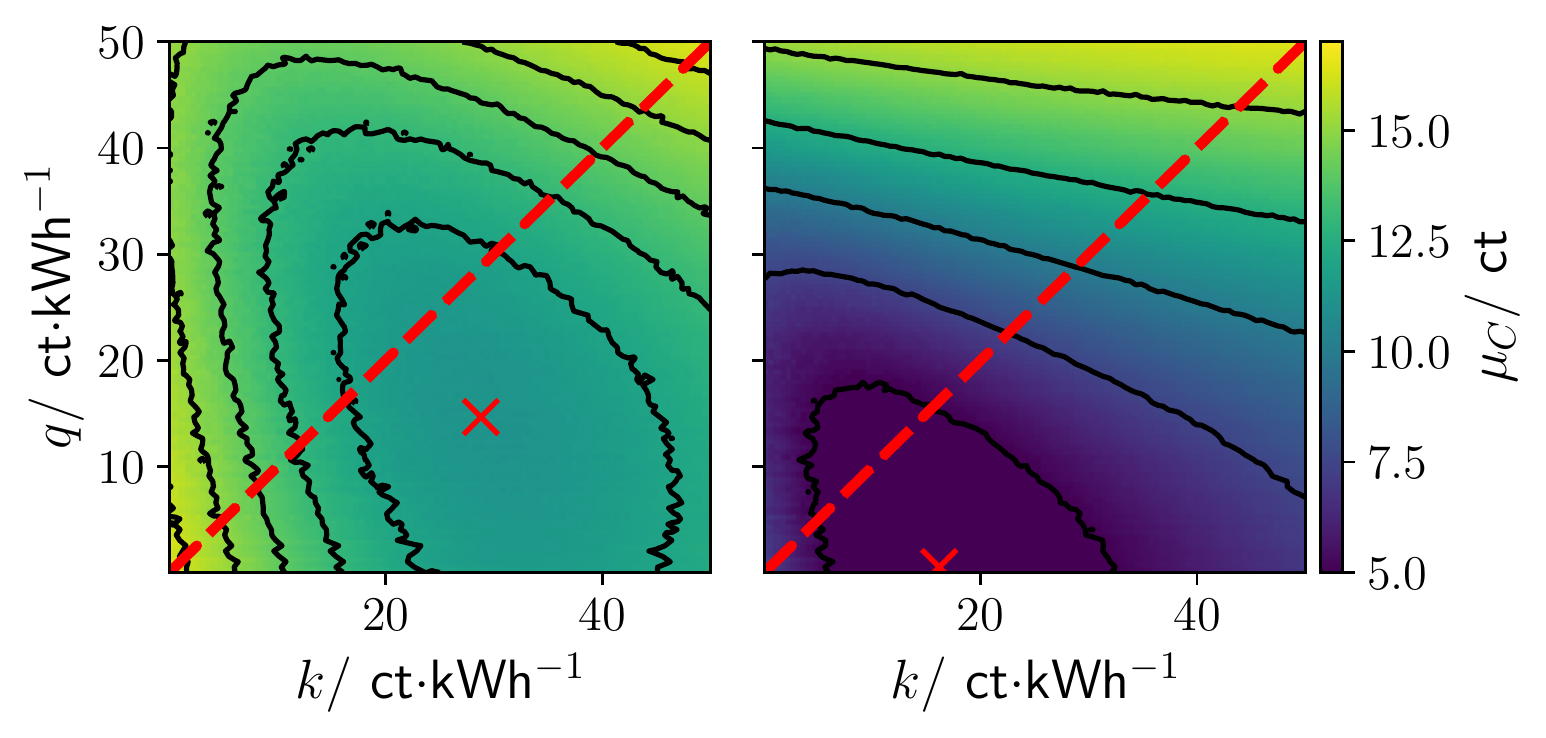}
    \caption{
    Reduction of the electricity costs of a single household by a DR/BESS system.
    We plot the average electricity costs $\mu_C$ as a function of the control system parameters $q$ and $k$ for two values of the BESS capacity: $S_{\rm Cap} = 10$kWh (left) and $S_{\rm Cap} = 40$kWh (right) and a charging rate of $c_r = .5$ h$^{-1}$.
    For the larger storage sizes $S_{\rm Cap}=40$kWh, a reduction in $\mu_C$ by a factor of approximately $3$ is possible compared to a storage size of $S_{\rm Cap}=10$kWh.
    The red cross denotes the optimum choice of the parameters $q^*$ and $k^*$ for which $\mu_C$ assumes its minimum, while the red dashed lines indicate the line for the constant strategy as explored in Fig.~\ref{fig:constant_law}.
    }
\label{fig:strategy_scan}
\end{figure}

We now turn back to the original control law given in Eq.~\eqref{eq:linear_law}, where the controller takes into account the state of charge of the battery. The control law is characterized by two parameters, $k$ and $q$, which are chosen to minimize the average costs $\mu_C$. In particular, we carry out a parameter scan for any given BESS system to find the optimum values $q^*$ and $k^*$, as shown in Fig.~\ref{fig:strategy_scan}. 
We find that a household can reduce its electricity costs considerably by the DR system depending on the size of the BESS. For a BESS capacity of $S_{\rm Cap} = 40$~kWh, we find a reduction of $\mu_C$ by more than a factor of 4 at optimum parameters. In the following simulations, we will always assume that all households set the control parameters to the optimum values $k^*$ and $q^*$.

\begin{figure}
    \includegraphics[width=.9\columnwidth]{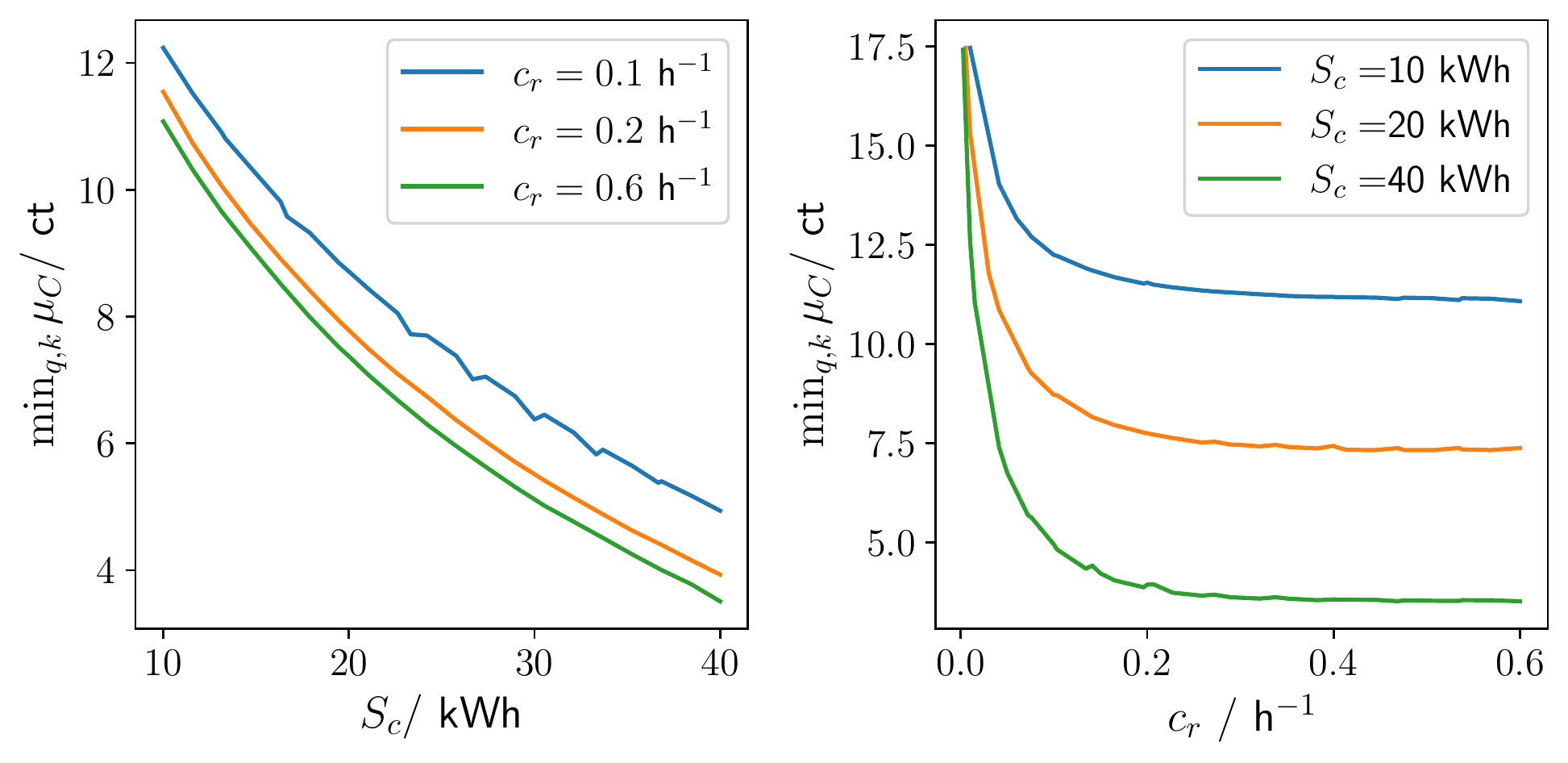}
    \caption{
    Reduction of average electricity costs $\mu_C$ for a single household as a function of the capacity $S_{\rm Cap}$ and charging rate $c_r$ of the BESS. In all cases, the respective optimal control parameters $k^*$ and $q^*$ were used.
    The minimal cost decrease monotonically with both $S_{\rm Cap}$ and the charging rate $c_r$, where the latter shows a pronounced saturation for $c_r \ge 0.2$ h$^{-1}$.    
    }
    \label{fig:prize_vs_scap_cr}
\end{figure}

A systematic study of the impact of the technical parameters of the BESS on the DR effect is provided in Fig.~\ref{fig:prize_vs_scap_cr}. We find that the average electricity cost $\mu_C$ at optimum parameter choices decreases monotonically with the available storage capacity $S_{\rm Cap}$. That is, the larger the BESS, the more it can contribute to load shifting and hence to a reduction of household electricity cost. The slope decreases slightly with the capacity $S_{\rm Cap}$, but we see no pronounced saturation effect for values up to $S_{\rm Cap} = 40$kWh considered in our simulations. For a fixed storage capacity $S_{\rm Cap}$, the average price drops rapidly with the maximum charging rate $c_r \, S_{\rm Cap}$ until it saturates at $c_r \approx 0.2$h$^{-1}$.

\section{Systemic effects and Statistics of DR}

The result of the previous section confirms that DR can lead to a substantial reduction of a household's electricity costs by shifting electricity purchases to time intervals with lower prices. As low prices typically correspond to periods of high renewable power generation, this is considered beneficial for the operation and stability of the entire power system. We will now demonstrate an important limitation to this general conclusion due to the collective effects induced by real-time DR. 

To quantify the collective effects and the impact on the system, we simulate the operation of many households. All households $j=1,\ldots,N$ have different demand patterns $D_j(t)$ but react to the same price signal $p(t)$. For the sake of simplicity, we furthermore assume that the parameters of the BESS are identical and that each household chooses the same optimal control parameters $p^*$ and $k^*$. The impact on the electricity system is analyzed in terms of (i) the statistics of the total grid load $E_{\rm tot}(t) = \sum_{j=1}^N E_j(t)$ and (ii) the fraction of electricity purchased at a certain price $p$. The latter quantity is estimated from the simulation results as
\begin{equation}
   Z(p) =  \mathcal{N}^{-1}
       \sum_{t: p(t) \in [p,p+\Delta p]} 
       \sum_{j=1}^N  E_j(t) , 
       \label{eq:paid-prices-PDF}
\end{equation}
Here, the sum over the variable $t$ is restricted to time steps where the price satisfies $p(t) \in [p,p+\Delta p]$, i.e.,~where it falls in a small interval around the given price $p$. The variable $\mathcal{N}$ denotes a normalization constant which ensures that the integral over $Z(p)$ equals one such that we can interpret $Z(p)$ as the density of purchases at a certain price. 

\begin{figure}
    \includegraphics[width=0.9\columnwidth]{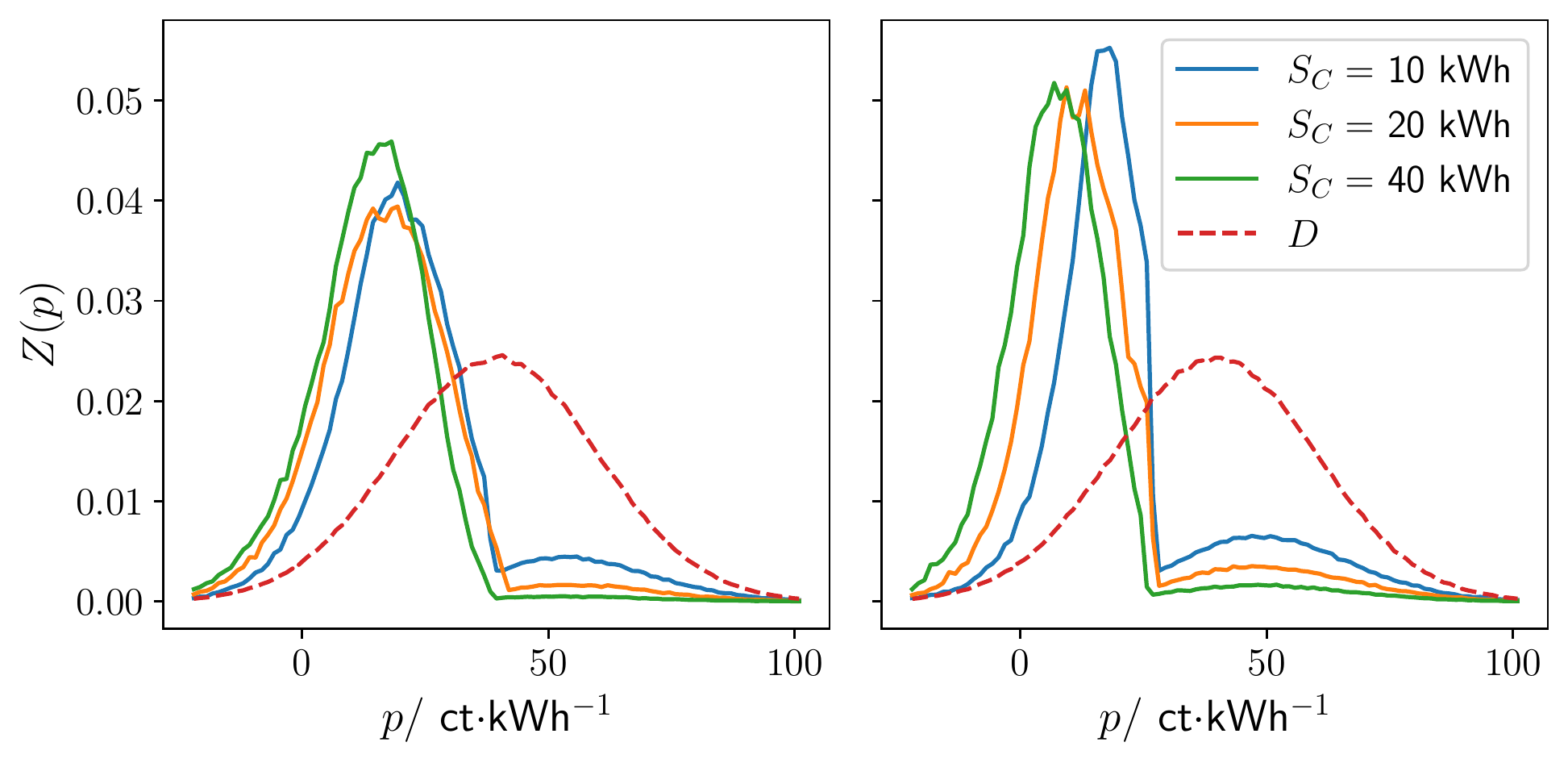}
    \caption{
    The likeliness of prices $p$ paid by the households. 
    The figure shows the density function defined in Eq.~\ref{eq:paid-prices-PDF} for DR systems with different storage capacities $S_{\rm Cap}$. In the absence of DR ($S_{\rm Cap} = 0$ kWh, dashed line), the density $Z(p)$ equals the density of the price time series $p(t)$. 
    In the presence of DR ($S_{\rm Cap} > 0$ kWh, solid lines), customers can shift the purchases to periods with lower prices. Hence, the density function $Z(p)$ is strongly shifted to lower values of $p$.
    In all cases, we use optimized parameters $k^*$ and $q^*$ for the controller. The charging flow to the batteries $c_r\cdot S_{\rm Cap}$ was chosen as $2$kW and $6$kW for the results presented on the left and right sides, respectively.
    }
\label{fig:price_density}
\end{figure}

\begin{figure}
    \includegraphics[width=.9\columnwidth]{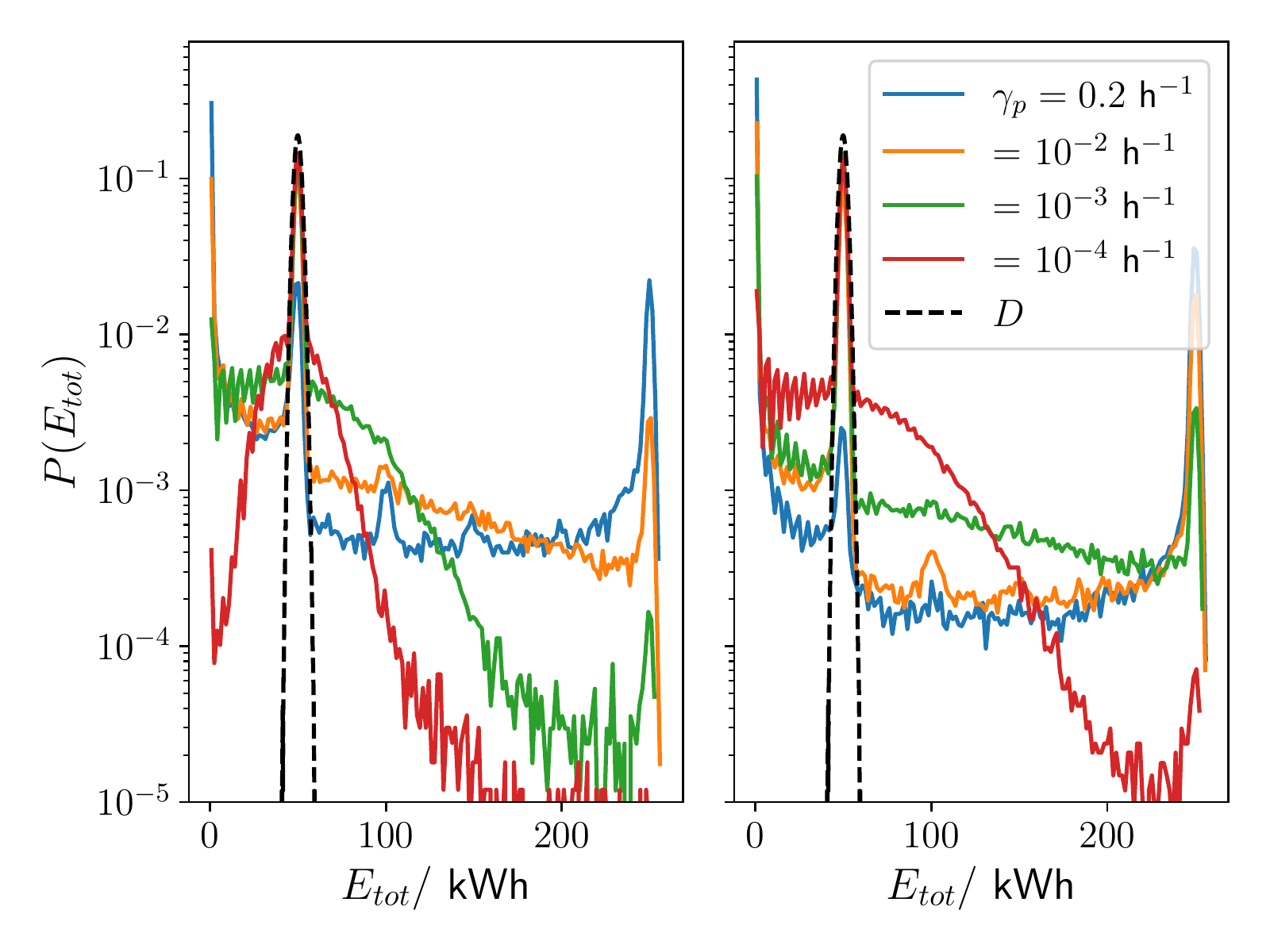}
\caption{
Distribution of total grid load $E_{tot}$ for different mean reversion rates $\gamma_p$ of the Ornstein-Uhlenbeck process giving the price.
The storage size of $S_{\rm Cap}=10$~kWh and $S_{\rm Cap}=40$~kWh are compared on the left and right, respectively.
In both cases, the total charging $c_r \cdot S_{\rm Cap}$ is chosen as $2$~kWh per hour. 
Black dashed line gives the distribution if no storage device would be used, which is equivalent to the distribution of the demand $D$.
When the time spent in either high or low price regimes is short enough to allow the battery device to be used effectively, the distribution of the total purchased energy $E_{tot}$, i.e., the stress to the grid, is broad, and situations with large total demand become very likely.
As the price dynamics gets slower, the distribution changes from an almost horizontal shape by narrowing considerably.
}
\label{fig:slow_prices}
\end{figure}

Consider first the case of no DR, which is recovered in the above model by setting $S_{\rm Cap} = 0$~kWh.
Electricity is drawn from the grid whenever demanded, $E_j(t) = D_j(t)$, independent of the actual price $p(t)$. Hence, the likeliness of buying at a certain price, $Z(p)$, equals the PDF of the market price $p(t)$, see Fig.~\ref{fig:price_density}. 
The individual purchases $E_j(t)$ fluctuate strongly, but the total system load $E_{\rm tot}(t)$ does \emph{not}. In fact, the individual fluctuations average out such that the total grid load is almost constant at a level of
\begin{equation}
    E_{\rm tot} \approx N \left\langle D_j(t) \right\rangle_{j,t} , 
\end{equation}
where the brackets denote averaging over time steps and households. The residual small fluctuations around this value are well described by a narrow Gaussian PDF, see Fig.~\ref{fig:slow_prices}. According to the central limit theorem, the relative width of the Gaussian decreases as $1/\sqrt{N}$.

This picture is completely altered in the presence of real-time DR. Customers shift their load to periods with lower prices to reduce their costs. Hence, the density function $Z(p)$ of purchases in a certain price interval is strongly shifted to lower values of $p$, as shown in Fig.~\ref{fig:price_density}.
Purchases during high-price time intervals are suppressed.
The larger the size of the battery $S_{\rm Cap}$ is,the less likely purchases at times with high price become, but they still occur occasionally.

In principle, load shifting is the desired effect of DR. However, the effects at different households are not independent but synchronized due to the coupling to the common price signal $p(t)$.
Consequently, the fluctuations at different households no longer average out, and the central limit theorem no longer applies.
The impact on the statistics of the total grid $E_{\rm tot}(t)$ is dramatic, as shown in Fig.~\ref{fig:slow_prices}.
Instead of a narrow normal distribution, we now find a wide bathtub-shaped distribution.
Events where all customers synchronously draw the maximum amount of power are quite likely. In particular, such events take place \emph{after} a longer period of high prices, where all BESS are empty, the acceptable prices $p_{a,j}$ are high, and all households start charging when the price finally drops \cite{krause2015econophysics}. These crucial events result in a peak of the distribution at the right edge at
\begin{equation}
    E_{\rm tot} \approx N \left( c_r S_{\rm Cap} + \left\langle D_j(t) \right\rangle_{j,t}\right) \Delta t,
    \label{eq:total-load}
\end{equation}
increasing linearly with the system size $N$.
Such periods with large energy purchases, i.e. stress to the grid, can be critical for the stability of the electric power grid by triggering malfunctions that might ultimately results in a blackout \cite{pourbeik2006anatomy}.
Remarkably, such events can not even be considered rare, as the probability density shows pronounced peaks.
We note that similar distributions with peaks at the right edge have been intensively studied in reliability theory \cite{sornette2009dragon}. 
In Fig.~\ref{fig:slow_prices}, it is also shown how this effect changes for different mean reversion rates of the price $\gamma_p$. 
If smaller and smaller $\gamma_p$ are used to generate the price time series $p(t)$, high stress situations become more and more unlikely since the battery storage device cannot sustain the long periods of high prices and the demand response effect is diminished.
Although one might consider these situations as preferable due to the absence of high stress situations, they are not beneficial to the individual households.
Since they are not able to escape high prices with the help of the BESS, the average cost of a household is considerably higher than in case with faster price dynamics, which is not desirable to individual households.

\begin{figure}
\includegraphics[width=.9\columnwidth]{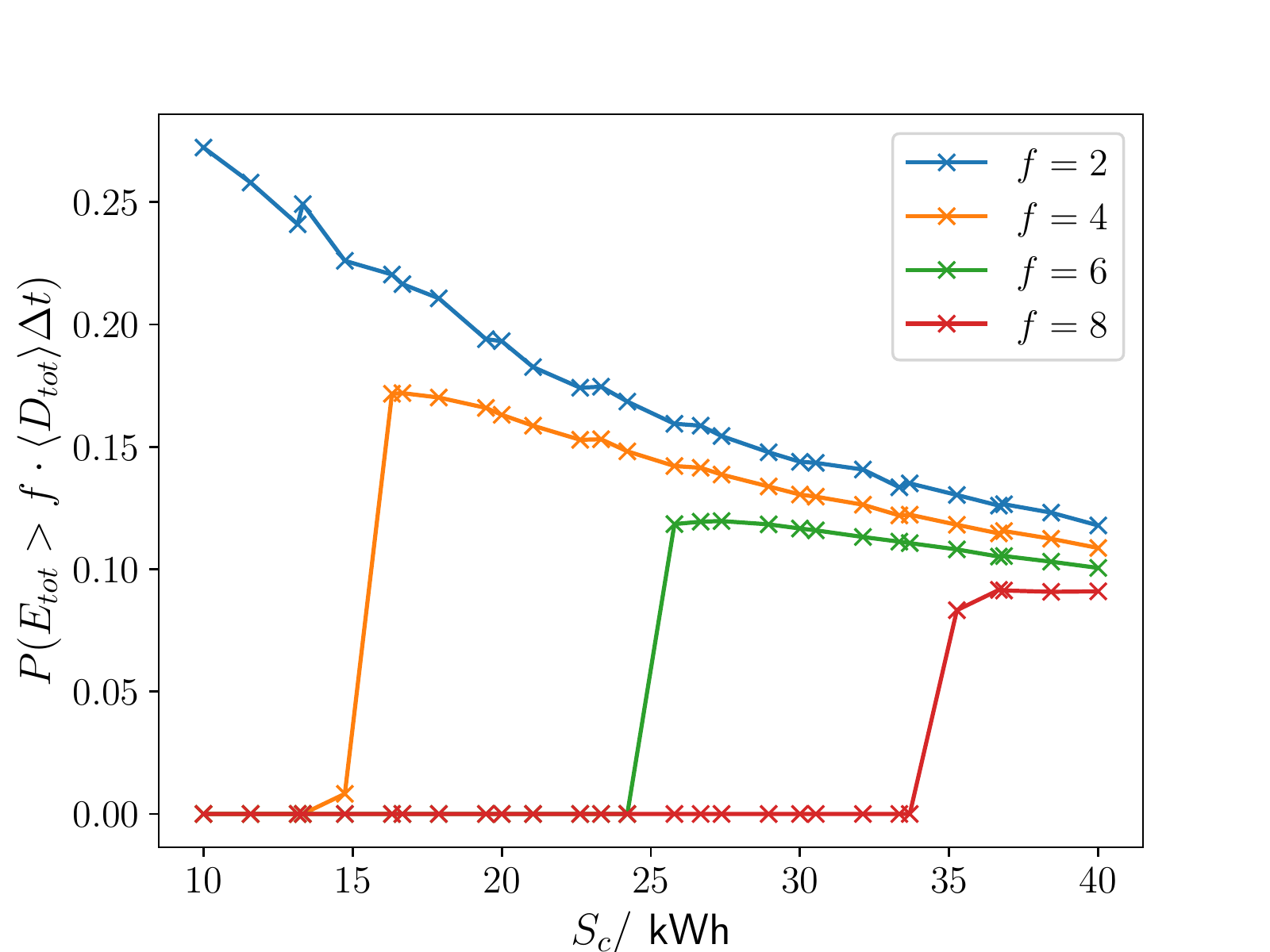}
\caption{Stress to the grid for different battery sizes.
The probability $P(E_{tot} > f \cdot \langle D_{\rm tot} \rangle \cdot \Delta t) $ of the total purchases energy is larger than $f$-times the average total demand $D_{tot}$ at the optimal strategy parameters $q^*$ and $k^*$.
Using larger batteries by increasing their capacity $S_{\rm Cap}$ generally decreases the stress to the grid but for one important effect. 
There is a sharp increase in the likelihood of very high stress situations for different $f$ values if one increases the capacity of the batteries.
}
\label{fig:high_purchase_results}
\end{figure}

To further quantify the likelihood of situations that strongly affect the grid, we evaluate the probability $P(E_{tot} > f \cdot \langle D \rangle \cdot \Delta t)$ that the purchases energy exceeds the average demanded energy $\langle D \rangle \cdot \Delta t$ by a factor of $f$. Results are shown as a function of the capacity $S_{\rm Cap}$ of the BESS in Fig.~\ref{fig:high_purchase_results}.
Without DR, extreme events with $f>2$ are never observed in our simulations. This is a direct consequence of the central limit theorem, which states that large deviations from the mean are exponentially unlikely.
DR now makes these events possible as the demand is accumulated during time periods with low prices. In particular, we find that extreme events become possible if the capacity $S_{\rm Cap}$ exceeds a threshold value. If the capacity increases further, the likeliness decreases again because purchases are further concentrated to fewer and fewer points in time. That is, extreme events become less likely but more pronounced in their magnitude.

\section{Discussion}

Demand response (DR) summarizes a variety of approaches to adapt the demand for electric power to better match the supply. This can be achieved by load shifting -- consumers shift their demand in time and may receive financial compensation for the utility company in return. DR can be an important source of flexibility in future renewable power systems, where the generation is volatile and cannot be easily adapted to the demand.

In this article, we have analyzed a model demand response system from a statistical viewpoint. Load shifting is realized by the optimized charging of a household battery electric storage system in response to real-time electricity pricing. Such storage systems are often installed together with a rooftop photovoltaic system, and it becomes increasingly important to their role in the operation of the entire system. 

On average, the model DR systems provide the desired load shifting effect. However, the statistics of the grid loads change dramatically, which may have unwanted or even harmful effects. These effects manifest the collective behaviour of many DR systems driven by the same price signals. Without DR, the electricity load of single households is largely uncorrelated besides the common daily profile. Hence individual fluctuations average out, and the total grid load is smoothed. With DR, the electricity load can get synchronized. The smoothing effect is lost, and we observe pronounced peaks instead. The distribution of the grid load then assumes a bathtub shape with pronounced peaks at zero and peak load.

Importantly, the demand peaks do not necessarily occur during the periods of the lowest prices. Instead, they may also occur if the price drops after a long period of high values. In such a case, DR operation may be counter-productive for system stability, introducing demand peaks at times of limited generation. 

In conclusion, we have demonstrated that demand response may induce load shifting patterns with intricate statistical properties. An analysis of the potentials of DR should not neglect the complexity of the emergent statistics. A comprehensive assessment of the benefits of DR requires a comprehensive statistical analysis of the systemic impacts.

\section*{Acknowledgements}

We gratefully acknowledge support from the German Federal Ministry of Education and Research (BMBF) via the grant \textit{CoNDyNet} with grant no. 03EK3055 and the Helmholtz Association via the grant \textit{Uncertainty Quantification -- From Data to Reliable Knowledge (UQ)} with grant no.~ZT-I-0029.
We also thank Benjamin Sch\"afer for stimulating discussions.

\FloatBarrier
\bibliography{ref}

\end{document}